\documentclass[12pt]{article}
\usepackage{amsmath}
\usepackage{amssymb}

\usepackage{epsf,graphicx,rotating,color}
\usepackage{bm,bbm}

\newcommand{\rme}{{\rm e}}
\newcommand{\rmi}{{\rm i}}

\title{Simulation of static and random errors on Grover's search algorithm 
  implemented in a Ising nuclear spin chain quantum computer with few qubits}
\author{T. Gorin, L. Lara and G.V. L\'opez\\ 
\small Departamento de F\'{\i}sica, Universidad de Guadalajara\\
\small Blvd. Marcelino Garc\'{\i}a Barragan y Calzada Ol\'{\i}mpica\\
\small 44840 Guadalajara, Jalisco, M\'exico}

\begin{document}

\maketitle

\begin{abstract}
We consider Grover's search algorithm on a model quantum computer implemented 
on a chain of four or five nuclear spins with first and second neighbor 
Ising interactions. Noise is introduced into the system in terms of random
fluctuations of the external fields. By averaging over many repetitions of the
algorithm, the output state becomes effectively a mixed state. We study its 
overlap with the nominal output state of the algorithm, which is called 
fidelity. We find either an exponential or a Gaussian decay for the
fidelity as a function of the strength of the noise, depending on the type of
noise (static or random) and whether error supression is applied
(the $2\pi k$-method) or not.
\end{abstract}

PACS: 03.67.-a,3.67.Lx, 03.65.Ta, 03.67.Dd, 03.67.Hk

\section{\label{I} Introduction}

It is well known that Grover's search algorithm for searching an item in a 
quantum data base of length $N=2^n$, where $n$ is the length of the quantum 
register made up of n-qubits, provides a quadratic speed up with respect any 
classical search algorithm~\cite{Gro1,Gro2}. Some applications of this 
algorithm have already been done~\cite{Ter,Bra,Cerf,Gin,Car,Yang}, and its 
physical implementation has been made for few 
qubits~\cite{Chu,Jon,Kwi,Lon01PLA}. Numerical simulations of this algorithm 
have also been made for a 4-qubits solid state 
quantum computer which is made up of a chain of four Ising paramagnetic nucleus 
of spin one-half interacting with the first and second spins 
neighbors~\cite{Lop1}. This simulations use RF-pulses to implement quantum 
gates~\cite{LoLa1}, and the $2\pi k$-method~\cite{Ber1} to control the non 
resonant transitions in the system. At the moment, there has not been a 
physical realization of this quantum computer. However, since the full 
Hamiltonian can be explicitly written for this system, several useful 
analytical and numerical studies have been made~\cite{Ber2,Lop2} where the 
results can be extrapolated to other solid state spins quantum computers. On 
the other hand, one important aspect of the simulation or implementation of a 
quantum algorithm is the effect that the  noise can cause to the performance of 
this algorithm which may arise from error in the quantum parameters or from 
the interaction with the environment, or both. This noise may disrupt the 
algorithm (decoherence) and, eventually, destroy it. Previous studies of the
effect of noise on a realisation of Grover's quantum search algorithm have
been based on description in terms of abstract gates, which have been turned
into noisy gates~\cite{Pab,LoLi,Ell,Son,Sal08} and/or the introduction of 
dissipation and decoherence within the Born-Markov 
approximation~\cite{ZhiShe06}. By contrast, the present model is based on a 
definite quantum system, with a specific Hamiltonian, where unitary errors and 
noise have a clear physical interpretation. In our model, unitary errors arise
from non-resonant transitions, whereas noise arises from fluctuations in the
external fields -- either the constant magnetic field or the radio-frequency 
pulses. 

In a previous paper~\cite{Lop1}, we studied unitary errors in the realisation
of Grover's quantum search algorithm on our quantum computer model. The present
work is devoted to the study of noise in essentially the same setting. In
Sec.~\ref{A}, we introduce the physical model quantum computer. In 
Sec.~\ref{G}, we explain how the quantum search algorithm is implemented with
the help of certain radio-frequency (RF-) pulses. Sec.~\ref{H} describes the
implementation of noise, by letting the Rabi frequency (Rabi-frequency noise),
or the Larmor frequencies (Larmor-frequency noise) fluctuate. The effect of
noise on the quantum computation is quantified using the fidelity, first 
introduced in~\cite{Per} in an attempt to transfer the concept of exponential
instability from classical to quantum mechanics. For a recent review on 
fidelity, see~\cite{GPSZ06}. Sec.~\ref{N} presents our actual
numerical results, and we summarize our findings in Sec.~\ref{C}.

\section{\label{A} Hamiltonian of the model quantum computer}

The model of the system can be found in reference~\cite{Ber1}, and the explicit 
Hamiltonian for simulation of the  Grover' search algorithm is given 
by~\cite{Lop2}
\begin{equation}
H= H_0 + W(t) \; ,\qquad
\label{A:defH}\end{equation}
where one has that
\begin{equation}
\frac{1}{\hbar}\; H_0= -\sum_{k=0}^3 w_k\; I^z_k 
 - 2J\sum_{k=0}^2 I^z_k\, I^z_{k+1} - 2J'\sum_{k=0}^1 I^z_k\, I^z_{k+2} \; , 
\label{A:defHo}\end{equation}
and
\begin{equation}
 W(t)= -\, \frac{\hbar\Omega}{2}\sum_{k=0}^{3}\left( 
   \rme^{\rmi (wt+\varphi)} I_k^+ + \rme^{-\rmi (wt+\varphi)} I_k^-\right) \; ,
\label{A:defW}\end{equation}
where $\omega_k$ is the Larmor frequency of the $k$'th-spin, $J$ and $J'$ are 
the coupling constant to first and second neighbor spins, $\Omega$ is the Rabi 
frequency, $\omega$ and $\varphi$ are the frequency and phase of the 
radio-frequency pulse. The operators $I_k^z$,  $I_k^+$  and $I_k^-$  are single 
qubit operators acting on the kth-qubit of the registers 
$\{|\alpha_3\alpha_2\alpha_1\alpha_0\rangle\}$ with $\alpha_i=0,1$ for 
$i=0,1,2,3$. These $2^4$-states are eigenstates of the 
Hamiltonian $H_0$, and the action of the operator $I_k^z$ on is associated 
qubit is as follows
\begin{equation}
I_k^z|\alpha_k\rangle=\frac{(-1)^{\alpha_k}}{2}|\alpha_k\rangle\ .
\label{I:Ik}\end{equation} 
The operator $I_k^+=(I_k^-)^{\dagger}=|0_k\rangle\langle 1_k|$%
\footnote{Note, in standard spin algebra, we would expect
$\hat I^+= \hat I^x + \rmi\, \hat I^y = |1\rangle\langle 0|$ being the ascend 
operator.}
represents the descend  operator for the qubit $|\alpha_k\rangle$, and the 
energy difference between two eigenstates of $H_0$ is
\begin{align}
E_{\alpha|\alpha_k=1} - E_{\alpha|\alpha_k=0} &= \hbar\big\{\, w_k 
   + J\, [ (-1)^{\alpha_{k+1}} + (-1)^{\alpha_{k-1}}]
   + J'\, [ (-1)^{\alpha_{k+2}} + (-1)^{\alpha_{k-2}}]\, \big\}\notag\\
&= \hbar\, \big (\, w_k + \mu J + \nu J'\, \big )\; , \qquad
\mu,\nu\in\{ -2,-1,0,1,2\}\; .
\end{align}
Any quantum gate is realized within the interaction picture, applying defined 
RF-pulses with rectangular shape, choosing the frequency $\omega$ in resonance 
with the appropriated transitions, and given the time ($\tau$) duration of the 
pulse. The corresponding evolution operator  for the 
whole quantum system, is denoted by $R_k^{\mu\nu}(\varphi,\theta)$, 
where $\theta=\Omega\tau$, and the RF-frequency is set equal to $w= w_k + \mu J + \nu J'$ (resonant situation).

\paragraph{Degeneracies in \boldmath $H_0$}
At a first glance, one would expect accidental degeneracies in the Hamiltonian
$H_0$ to cause problems in the realisation of any quantum algorithm, since a
given microwave field could drive unwanted transitions involving degenerated
states. Such degeneracies can be avoided requireing the Larmor frequencies
$w_k$ to increase in powers of two: $w_k \propto 2^k$. Without interactions
$J=J'=0$, this produces an equidistant spectrum; for $J'\ll J\ll w_0$ this 
guaranties a spectrum without any degeneracies. Such a choice of the Larmor 
frequencies would destroy the scalability of the present system as a model for 
quantum computation.

However, as it turns out degeneracies are not really a problem. This is due to 
the fact that the interaction operator couples states $|\alpha\rangle$ and 
$|\beta\rangle$ only if they differ in no more than one qubit. In the 
interaction picture, $W(t)$ becomes:
\begin{equation}
\tilde W_{\alpha\beta}(t)= -\, \frac{\hbar\Omega}{2}\begin{cases}
   \rme^{\rmi\, (\Delta_k t + \varphi)} &: 
      \beta-\alpha = (0,\ldots, 1_k, \ldots, 0)\\ 
   \rme^{-\rmi\, (\Delta_k t + \varphi)} &: 
      \alpha-\beta = (0,\ldots, 1_k, \ldots, 0)\\
   0 &: \text{else}\end{cases} \; ,  \qquad
\Delta_k = w - (w_k + \mu J + \nu J')\; ,
\end{equation}
where the notation $\alpha - \beta = (0,\ldots, 1_k, \ldots, 0)$ means that 
the multi-indices $\alpha,\beta$ are equal except at one position (qubit $k$),
where the qubit in the $\alpha$-state is excited and the qubit in the
$\beta$-state is not. For an accidental degeneracy caused by the unfortunate
choice of the Larmor frequencies, there must exist transitions such that
$w_k + \mu J + \nu J' = w_l + \mu' J + \nu' J'$ for some $k\ne l$. But this can
be savely avoided by choosing the spacing between neighboring Larmor 
frequencies much larger than the coupling constants: 
$|w_k - w_{k-1}| \ll J \ll J'$. Since this allows for a linear increase in the
Larmor frequency as a function of the number of qubits, the scalability of the 
quantum computer model is not affected.

\paragraph{Near resonant approximation} 
Assume that the frequency $w$ of the 
RF-pulse matches a particular transition so well that all other transitions
can be neglected. For a rectangular RF-pulse as described above, with detuning 
$\Delta= w - w_k -\mu J - \nu J'$, the evolution operator in the basis of the
two states involved read:
\begin{equation}
U_{\rm pulse}= \begin{pmatrix} \rme^{\rmi\Delta\tau/2} & 0\\ 
   0 & \rme^{-\rmi\tau\Delta/2} \end{pmatrix}
\begin{pmatrix} 
  \cos\frac{\Omega_\rme\tau}{2} - \frac{\rmi\Delta}{\Omega_\rme}\, 
     \sin\frac{\Omega_\rme\tau}{2} & 
  \frac{\rmi\Omega}{\Omega_\rme}\, \rme^{\rmi\varphi}\, 
     \sin\frac{\Omega_\rme\tau}{2}\\[5pt]
  \frac{\rmi\Omega}{\Omega_\rme}\, \rme^{-\rmi\varphi}\, 
     \sin\frac{\Omega_\rme\tau}{2} &
  \cos\frac{\Omega_\rme\tau}{2} + \frac{\rmi\Delta}{\Omega_\rme}\, 
     \sin\frac{\Omega_\rme\tau}{2} \end{pmatrix}\; ,
\label{A:U}\end{equation}
where $\Omega_\rme= \sqrt{\Omega^2 + \Delta^2}$. For $\Delta=0$, we obtain the
resonant transitions:
\begin{equation}
R(\varphi,\theta)= \begin{pmatrix}
   \cos\theta/2 & \rmi\, \rme^{\rmi\varphi}\, \sin\theta/2\\[5pt]
   \rmi\, \rme^{-\rmi\varphi}\, \sin\theta/2 & \cos\theta/2\end{pmatrix}\; ,
\qquad \theta= \Omega\tau\; .
\label{A:R}\end{equation}

\section{\label{G} Implementation of Grover's search algorithm}

For our case, Grover' search algorithm~\cite{NieChu00} requieres three qubits 
to prepare the "enquiry" states and a single qubit ("ancilla") required by the 
oracle to communicate its answer. Starting from the ground state, 
$|0_3 0_2 0_1 0_0\rangle$, a superposition state with the qubits $k=0,2,3$ is 
generated by applying the corresponding Hadamard gates, 
\begin{equation}
H^{(3)}=H_3 H_2 H_0\ ,
\label{I:H3}\end{equation}
to this state, where $H_k$ is the Hadamard gate acting to the kth-qubit. Then, 
one applies the Grover operator,
\begin{equation}
G=O_{\alpha}H^{(3)} S_0 H^{(3)}\ ,
\label{I:G}\end{equation}
twice, since for $ N=8=2^3 $ registers, the probability to find the searched 
state is maximum for $[\pi\sqrt{8}/4-1/2]\approx 2$  applications of Grover 
operator (steps). The operator $S_0$ represents the phase inversion of the 
searched state, and the operator $O_{\alpha}$ represents the oracle where 
$\alpha$ is the index (state) to be found in decimal notation. The 
implementation of these operatos by pulses is found in~\cite{Lop1},
\begin{equation}
H_0= \left\{ \textstyle 
   \prod_{\mu,\nu=-1,1} R_0^{\mu\nu}(\frac{\pi}{2},\pi) \right\}
   \left\{ \textstyle    
   \prod_{\mu,\nu=-1,1} R_0^{\mu\nu}(\frac{\pi}{2},\frac{\pi}{2}) \right\}
   \left\{ \textstyle    
   \prod_{\mu,\nu=-1,1} R_0^{\mu\nu}(\pi,\pi) \right\} \; .
\label{I:A0}\end{equation}
\begin{equation}
H_2= \left\{ \textstyle 
   \prod_{\mu,\nu=-2,0,2} R_2^{\mu\nu}(\frac{\pi}{2},\pi) \right\}
   \left\{ \textstyle    
   \prod_{\mu,\nu=-2,0,2} R_2^{\mu\nu}(\frac{\pi}{2},\frac{\pi}{2}) \right\}
   \left\{ \textstyle    
   \prod_{\mu,\nu=-1,1} R_2^{\mu\nu}(\pi,\pi) \right\} \; .
\label{I:A2}\end{equation}
\begin{equation}
H_3= \left\{ \textstyle 
   \prod_{\mu,\nu=-1,1} R_3^{\mu\nu}(\frac{\pi}{2},\pi) \right\}
   \left\{ \textstyle    
   \prod_{\mu,\nu=-1,1} R_3^{\mu\nu}(\frac{\pi}{2},\frac{\pi}{2}) \right\}
   \left\{ \textstyle    
   \prod_{\mu,\nu=-1,1} R_3^{\mu\nu}(\pi,\pi) \right\} \; .
\label{I:A3}\end{equation}
\begin{equation}
S_0=  R_1^{2,-1}(0,2\pi)\; R_1^{0,1}(0,2\pi)\; R_1^{0,-1}(0,2\pi)\;
   R_1^{-2,-1}(0,2\pi)\; R_1^{-2,1}(0,2\pi) \; ,
\label{I:S0}\end{equation}
\begin{alignat}{2}
O_0 &= R_1^{2,1}(0,2\pi)\qquad & O_8 &= R_1^{2,-1}(0,2\pi) \\
O_5 &= R_1^{-2,1}(0,2\pi)\qquad & O_{13} &= R_1^{-2,-1}(0,2\pi)\; .
\end{alignat}
One applies 70 pulses to represent the operator $H^{(3)}$ , for the 
superposition state,  and 146 pulses to represent  Grover operator to have a 
total of 362 pulses to implement Grover' search algorithm.

The duration of the RF-pulses used depends on the rotation angle $\theta$ and
on the Rabi frequency: $\tau_{\rm pulse}= \theta/\Omega$. Further below, we 
will measure time in units of $t_{\rm ph}= \pi/(2\Omega)$, the duration of a 
$\pi/2$-pulse.

\section{\label{H} Implementation of noise}

The Larmor frequencies $\omega_k$ for $k=0,1,2,3$ depend on a static magnetic 
field in $z$-direction with a strong field gradient:
$\omega_k=\gamma B_k$, where $B_k$ is the strength of the static 
magnetic field at the location of the $k$th-paramagnetic nucleus of spin 
one-half. The Rabi-frequency $\Omega$ is determined by the strength of 
the radio-frequency (RF-) pulses which is assumed to be constant during a 
pulse: $\Omega=\gamma B_t$, where $\gamma$ is the gyromagnetic ratio 
and $B_t$ is the amplitude of the transversal RF-pulse. The environment or 
imperfections in the magnetic field produce noise  which, in turn, will produce 
variations in the above parameters. To study the effect of these variations, we 
consider random detunings of the form
\begin{equation}
\omega_k=\omega_{k{\rm o}}+\epsilon_L\, \xi
\label{I:wko}\end{equation}
and
\begin{equation}
\Omega=\Omega_{\rm o} + \epsilon_R\, \xi\; ,
\label{I:Oo}\end{equation}
where $\omega_{k{\rm o}}$ for $k=0,1,2,3$ and $\Omega_{\rm o}$ are the Larmor 
and Rabi frequencies without noise, $\xi$ is  a random Gaussian variable,
centered at zero, with unit variance. The parameters $\epsilon_L$ 
and $\epsilon_R$ determine the amplitude of the noise.

The two types of noise produce different types of errors as can be seen from
the evolution operators in the resonant approximation, Eqs.~(\ref{A:U}) 
and~(\ref{A:R}). In the case of noise on the Rabi frequency, the evolution
operator remains of the form~(\ref{A:R}), only the angle $\theta$ suffers a
change:
\begin{equation}
\theta \to \theta'= \theta + \Delta_R\; , \qquad
\Delta_R= \epsilon_R\xi\, \tau\; ,
\end{equation}
where $\tau$ is the duration of the RF-pulse.
In the case of noise on the Larmor frequencies, the evolution operator gets 
distorted in a more complicated way: With $\theta= \Omega\tau$ and 
$\Delta_L= \epsilon_L\xi\, \tau$ we find:
\begin{equation}
R(\varphi,\theta) \;\to\; R'(\varphi,\theta)=
\begin{pmatrix} \rme^{\rmi\Delta_L/2} & 0\\ 
   0 & \rme^{-\rmi\Delta_L/2} \end{pmatrix}
\begin{pmatrix}
  \cos\theta/2 - \frac{\rmi\Delta_L}{\theta}\, \sin\theta/2 & 
  \rmi\, \rme^{\rmi\varphi}\, \sin\theta/2\\[5pt]
  \rmi\, \rme^{-\rmi\varphi}\, \sin\theta/2 &
  \cos\theta/2 + \frac{\rmi\Delta_L}{\theta}\, \sin\theta/2 \end{pmatrix}\; ,
\end{equation}
up to second order in $\Delta_L= \epsilon_L\, \xi\, \tau$. 
In both cases, the strength of noise must be compared to the smallest
energy (frequency) scale in the system. This is $J'$ the coupling between 
second nearest neighbors.

Besides the amplitude also the frequencies of the noise fluctuations are 
important. It makes a difference whether the variation of the Larmor 
frequencies (Rabi frequency) happen on a time scale which is of the order of 
the duration of the whole algorithm (low frequency noise) or of the duration of 
a single RF pulse. To investigate this issue, we consider two types of noise 
(errors): {\em Static noise}, where the respective frequencies are detuned for 
each realization, but kept fixed during the whole algorithm. {\em Random 
noise}, where the respective frequencies are detuned anew for each RF-pulse. 
The latter simulating noise which fluctuates on a time scale of the order of 
the inverse Rabi frequency.

\paragraph{Fidelity}
In order to measure the effect of noise on the realisation of Grover's quantum
search algorithm, we use the fidelity (or quantum Loschmidt 
echo)~\cite{Per,GPSZ06}. We perform averages over $n_{\rm rep}$ repetitions
(where $n_{\rm rep}$ is of the order of 100) of the algorithm. This effectively 
turns the final state of the quantum computer into a mixed state 
$\varrho_{\rm real}$. The fidelity is computed by projection of 
$\varrho_{\rm real}$ on the outcome of the ideal realisation of the algorithm:
\begin{equation}
F_{\rm end}= 
  \langle\Psi_{\rm ideal}|\; \varrho_{\rm real}\; |\Psi_{\rm ideal}\rangle\; ,
\qquad
\varrho_{\rm real}= \frac{1}{n}\sum_{r=1}^n\; 
   |\Psi_{\rm real}^{(r)}\rangle\langle\Psi_{\rm real}^{(r)}|\; ,
\label{N:av}\end{equation}
where $|\Psi_{\rm real}^{(r)}\rangle$ are the final states obtained from the
algorithm taking into account the random detuning of the respective 
frequencies.

When analysing the final fidelity, there are two options: we can compare
the final state after the evolution in the presence of noise and unwanted  
transitions with the final state obtained from an error-free execution of the 
quantum protocol, or we can compare the final state with the desired 
target state. These two options are not the same, because the Grover protocol 
arrives at the desired target state only with a final probability, which is 
close to but still different from one~\cite{Lon01PRA}. Both options yield 
very similar results for the behavior of fidelity. For the results presented in 
this paper, we chose the first option.

\section{\label{N} Numerical method and results}

We chose $J=10$ (one order of magnitude smaller than the spacing between the
Larmor frequencies) and $J'=0.4$ (one order of magnitude smaller than $J$). In 
this way, made sure that the dominant source for errors were the near resonant 
transitions with detunings of the order of $J'$. These parameters are all 
frequencies -- see Eq.~(\ref{A:defHo}) -- given in units of 
$2\pi\times {\rm MHz}$. Note however that since the model depends only on their 
relative values, we may also choose any other unit. Unless stated otherwise, we 
choose the Rabi frequency as $\Omega_o = 0.1008$ according to the 
$2\pi k$-condition, such that near resonant transitions due to the 
second-neighbor coupling are efficiently suppressed.

The evolution of the quantum state in the 
qubit-register is calculated numerically by solving the time-dependent 
Schr\" odinger equation in the interaction picture. For that we have to solve
a first order ordinary differential equation, where the state vector is 
represented by $N=2^n$ complex amplitudes. In view of the rapidly oscillating
right hand side (the interaction matrix), any sophisticated adaptive stepsize
routine is of limitted use. We therefore applied a simple 4'th order 
Runge-Kutta algorithm to solve the differential equation, where we fixed the 
stepsize $\Delta t$ according to the largest Larmor frequency $w_n$ which is
the largest frequency in the problem: $\Delta t= \pi/(2 w_n)$.

The computation time depends mainly on the time needed to evaluate the right 
hand side of the differential equation. We have to calculate the components of
an $N$-dimensional vector, and for each component we have to sum over the $n$
non-zero matrix elements of the interaction matrix. This means that the 
time needed for the evaluation of the right hand side scales as $nN= 2^{n+1}$ 
with the number of qubits. In addition, more qubits require a larger pulse 
sequence to apply the Grover algorithm. Ultimately, the length of the pulse
sequence would scale with $\sqrt{N}$ as explained for instance in 
Ref.~\cite{NieChu00}. This would mean that overall, the computation time scales
as $2^{3n/2}$ .
The simulation presented in this work are done on standard single processor
machines. With our hardware a single run of the Grover algorithm with 4 or 5
qubits takes approximately 4 minutes computing time. 

\paragraph{Choice of the Larmor frequencies}
On the basis of the discussion of accidental degeneracies in section~\ref{A}, 
we expect that the Larmor frequencies may be chosen to increase linearly with
the number of qubits. Thus, we normally set
\begin{equation}
w_0= 50\; , \quad w_1= 200\; , \quad w_2= 350\; , \quad w_3= 500
\label{N:LarmorSet}\end{equation}
for the calculations with four qubits and in addition $w_4=650$ for the 5-qubit
calculations. However, before settling for these values we compared the errors
in different schemes (exponential increase, avoiding any degeneracies in $H_0$,
quadratic increase, and the present linear increase). We found consistently,
that the different schemes only lead to apreciable errors if at least one of 
the differences $w_k - w_l$ gets close to the energy scale of the coupling 
constants. Our choice in Eq.~(\ref{N:LarmorSet}) is perfectly save in view of 
the values chosen for $J$ and $J'$.

\subsection{Larmor frequency noise}

\begin{figure}[t]
\includegraphics[scale=0.7]{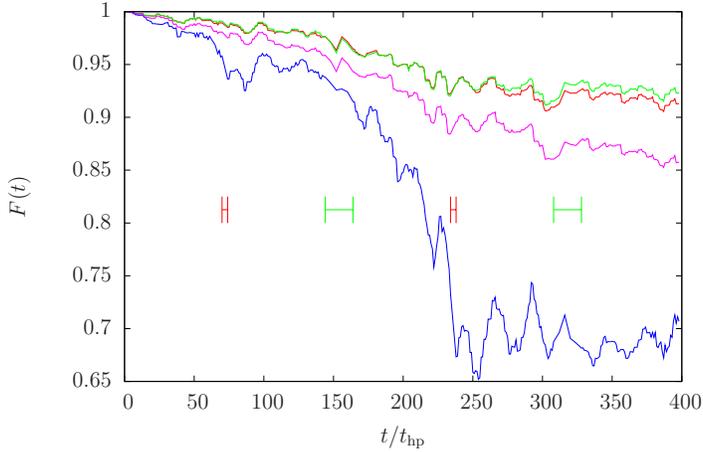}
\caption{Fidelity as a function of time for single runs of Grover's quantum 
search algorithm. Numerical simulations: without noise and exact (red line),
using the near resonant approximation (green line); with static Larmor 
frequency noise ($\epsilon_L= 0.02$) and exact (blue line), using the near 
resonant approximation (pink line). The red and green vertical bars indicate 
the position of pulses implementing the oracle (red bars) and the conditional 
reflection (green bars). Time is indicated in units of $t_{\rm hp}$ the 
duration of a $\pi/2$-pulse.}
\label{f:LarmorSingle}\end{figure}

To start our study, we analyse single runs without any averaging. We compare
calculations with and without noise, as well as exact calculations and 
calculations which apply the near resonant approximation. We use a quantum
register of 3+1 qubits to simulate Grover's search algorithm. The qubits are
arranged according to the string {\tt xx.x}, where x represents a data qubit,
while the . represents the auxiliary qubit which is used to implement the 
oracle and the conditional reflection. The algorithm consists of an initial 
sequence of Hadamard gates $H^{(3)}$ followed by two Grover steps. 

In the simulation for Fig.~\ref{f:LarmorSingle} the algorithm searches for the 
state 0 (string {\tt 00.0}). On the graph we indicate the time intervals where 
the quantum oracle is implemented with red bars. These intervals therefore mark 
the start of a Grover step. Furthermore, we mark the time intervals where the 
conditional reflection is implemented with green bars. The red solid line shows 
the fidelity as a function of time for the exact calculation without noise. 
During the algorithm, the fidelity drops with more or less constant rate from 
$F(0)=1$ to $F_{\rm end} \approx 0.92$. This reminds us that the implementation 
of the quantum gates via RF-pulses as such already introduces errors. On the 
graph, time is indicated in units of $t_{\rm hp}$ the duration of a 
$\pi/2$-pulse (see end of Sec.~\ref{G}). The green solid line shows the 
same calculation, but with the near resonant approximation being applied. 
It follows the exact calculation very well. The near resonant approximation 
neglects the far resonant, where the frequency missmatch is of the order of the
Larmor frequencies. The fidelity is generally a bit larger than for the exact 
calculation. The agreement between the exact calculation and the 
near resonant approximation is completely lost as soon as the Larmor frequency 
noise (here it is static noise may be better described as random detunings)
is switched on (blue and pink solid lines). The blue line shows the exact 
calculation for $\epsilon_L= 0.02$ which only initially follows the curve for 
the fidelity decay without noise. More or less in the middle of the algorithm, 
the fidelity curve drops rather abruptly to a level of $F(t) \approx 0.7$, 
where it saturates. The near resonant approximation doesn't show this behavior. 
Instead it stays close to the curve which shows the fidelity decay without 
noise. We may conclude that the noise introduces additional errors primarily 
via such indirect transitions which are neglected in the near resonant 
approximation. In the simulations shown, there is no indication that the 
oracle or the conditional reflection play any special role in the loss of 
fidelity.

\begin{figure}[t]
\includegraphics[scale=0.7]{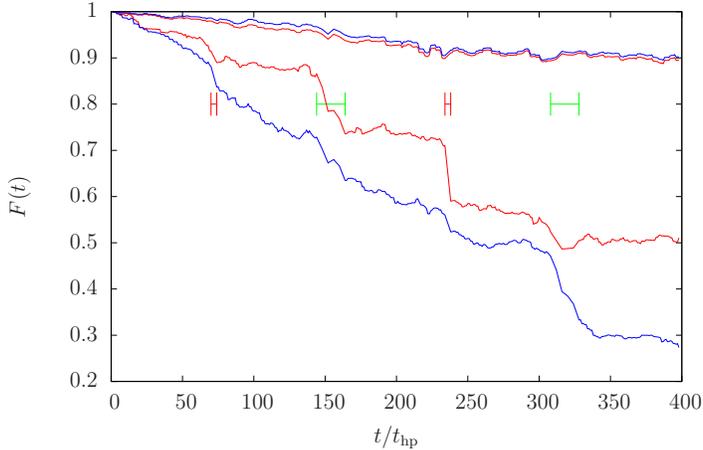}
\caption{The average fidelity as a function of time in the case of static (red
lines) and random (blue lines) Larmor frequency noise. Exact numerical
simulations and the near resonant approximations are shown in the same color. 
The latter largely over-estimate the fidelity and fail to reproduce its true 
behaviour. Red and green vertical bars again indicate those time intervals, 
where the oracle and the conditional reflections are applied.}
\label{f:LarmorAverage}\end{figure}

Figure~\ref{f:LarmorAverage} shows the average fidelity as described by 
Eq.~(\ref{N:av}) for static and random noise on the Larmor frequencies. The
simulations are averaged over an ensemble of $n_{\rm rep}$ independent 
repetitions of the algorithm. In the case of the exact numerical simulations,
$n_{\rm rep}= 50$ (static noise) and $25$ (random noise). In the case of the 
near resonant approximations, $n_{\rm rep}= 1000$ in both cases. This figure 
serves to demonstrate that the behavior of the fidelity may be very different 
for different repetitions -- at least for static noise (compare with 
Fig.~\ref{f:LarmorSingle}). We expect smaller variations within individual 
repetitions in the case of random noise, since there, the fidelity depends 
on a large number of random variables which tend to wash out any kind of 
extreme behavior. Interestingly, the simulations reveal that for static noise,
fidelity is lost to a large extent during the oracle and the conditional 
reflection pulses. For random noise, the fidelity drop is more uniform.
Finally, Fig.~\ref{f:LarmorAverage} again demonstrates the failure of the near
resonant approximation in the presence of Larmor frequency noise.

\subsubsection*{Static vs. random Larmor frequency noise}

Here, we analyse the average fidelity for the whole algorithm as a function of 
the noise amplitude. Averages are now taken over ensembles which contain 
typically about $n_{\rm rep}= 100$ repetitions. The statistical error (due to
the finite size of the ensemble) is estimated as follows: We divide the 
$n_{\rm p}$ fidelity values for each repetition into $p$ sub-ensembles of equal
size. This yields average fidelities $F_i,\, i=1,\ldots,p$. We take care that 
these sub-ensembles are sufficiently large such that the group-averaged 
fidelities are Gaussian distributed. That allows us to estimate the variance
of the full average: $F= p^{-1}\sum_{i=1}^p F_i$ by calculating the best 
estimate for the variance of the group averages:
\begin{equation}
{\rm var}(F) = \frac{1}{p}\; {\rm var}(F_i) \; , \qquad
{\rm var}(F_i) = \frac{1}{p-1} \sum_{i=1}^p (F_i - F)^2 \; .
\label{N:StatErr}\end{equation}
We then use the standard deviation $\sigma(F)= \sqrt{{\rm var}(F)}$ as an
estimate for our statistical error.

\begin{figure}[t]
\centerline{
\includegraphics[scale=0.8]{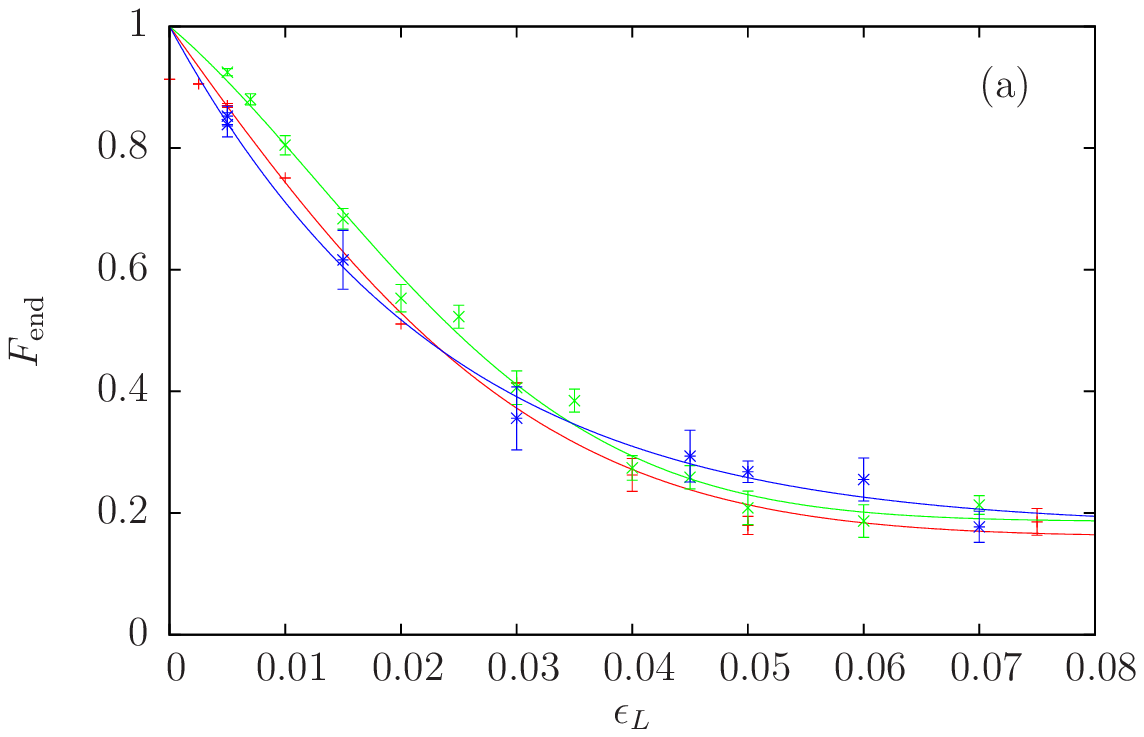}
\includegraphics[scale=0.8,clip=false,trim=1.3cm 0 0 0]{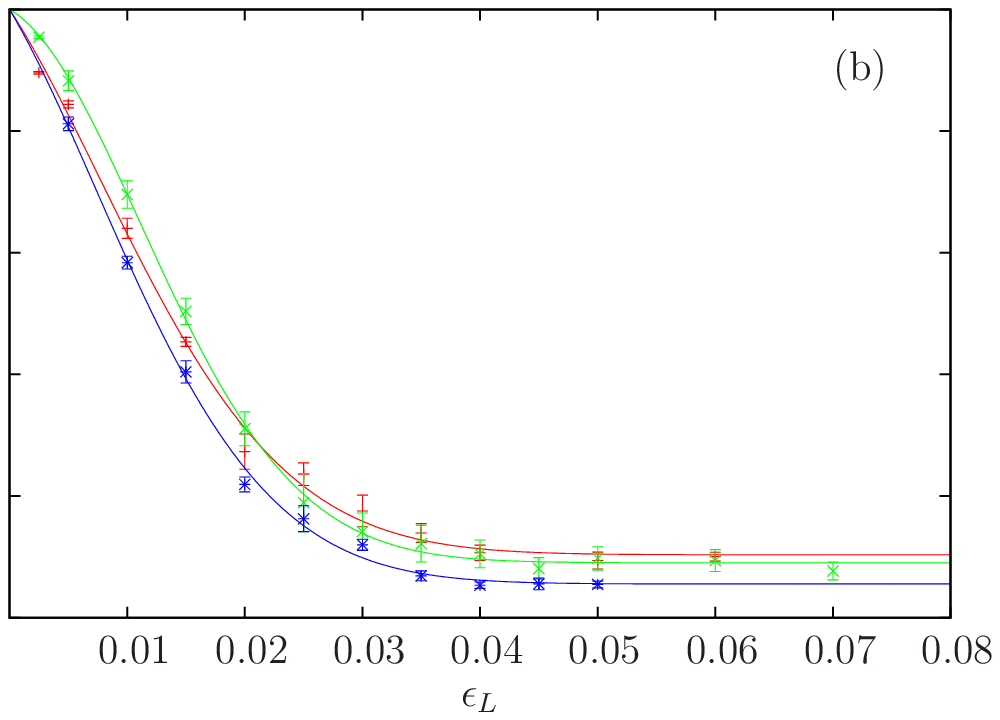}}
\caption{The fidelity $F_{\rm end}$ for static (a) and random (b) noise on the
Larmor frequencies. Exact numerical simulations: $4$ qubits, target state $0$ 
(red error bars); $4$ qubits, target state $8$ (green error bars); $5$ qubits,
target state $27$ (blue error bars). The solid lines of the corresponding
colour show a phenomenological fit with Eq.~(\ref{N:fit}) to the numerical 
data.}
\label{S:fig2}\end{figure}

Fig.~\ref{S:fig2} shows the final average fidelity $F_{\rm end}$ of the whole 
algorithm as a function of the amplitude $\epsilon_L$ of the Larmor frequency
noise. We compare the simulations with static noise (panel (a)) and random 
noise (panel (b)). In either case, we simulate the algorithm for four qubits
and target states $0$ and $8$, and also for five qubits and the target state
$27$. For a given noise model (static or random) the fidelity $F_{\rm end}$ 
behaves very similar in all cases. In panel (a) we observe a roughly 
exponential decrease for increasing $\epsilon_L$. At large values of 
$\epsilon_L$ the fidelity saturates at approximately 
$f_{\rm bas}\approx 0.175$. In panel (b), random noise, we observe a somewhat 
faster decay. Again the fidelity saturates for large $\epsilon_L$, but now the
fidelity saturates earlier than in the case of static noise. Also, the 
saturation level is somewhat smaller: $f_{\rm bas}\approx 0.08$. To compare 
the behavior of the final fidelity $F_{\rm end}$ more quantitatively, we 
consider the following phenomenological model function:
\begin{equation}
f(\epsilon)= f_{\rm bas} + (1-f_{\rm bas})\; \exp\left[ -  
   (\epsilon/\epsilon_2)^2 - \epsilon/\epsilon_1\right] \; ,
\label{N:fit}\end{equation}
with the fit parameters $f_{\rm bas}, \epsilon_2$, and $\epsilon_1$. In this 
parametrization, $\epsilon_2$ indicates the point where Gaussian decay becomes
important. Similarly, $\epsilon_1$ indicates the point where exponential decay
becomes important. For the three cases shown in Fig.~\ref{S:fig2}(a), static
noise, we obtain the following results:
\begin{equation}
\begin{array}{r|ccc}
                      & f_{\rm bas} & \epsilon_2 & \epsilon_1\\
\hline
\text{4 qubits, O(0)} & 0.161(49) & 0.046(20) & 0.0315(73) \\
\text{4 qubits, O(8)} & 0.186(19) & 0.0359(40) & 0.0510(80) \\
\text{5 qubits, O(27)} & 0.178(62) & 0.11(22) & 0.0235(23)\end{array} \; ,
\end{equation}
where the statistical uncertainties of the fit values are given in parenthesis
(they refer to the last two digits of the respective fit value). For the three 
cases shown in Fig.~\ref{S:fig2}(b), dynamic noise, we obtain:
\begin{equation}
\begin{array}{r|lll}
                      & f_{\rm bas} & \epsilon_2 & \epsilon_1\\
\hline
\text{4 qubits, O(0)} & 0.104(12) & 0.0224(20) & 0.0303(44) \\
\text{4 qubits, O(8)} & 0.0904(45) & 0.01870(55) & 0.083(16) \\
\text{5 qubits, O(27)} & 0.0555(83) & 0.0209(14) & 0.0291(34) \end{array}
\end{equation}
Comparing the reduction of the fidelity for
static and random noise, we find that random noise has a more destructive
effect. This holds even at very small noise levels. This result may be
surprising in view of opposite results in Refs.~\cite{LoLi,Frahm04}. However, 
note that both articles focus on larger quantum registers with more qubits
and correspondingly more Grover steps. This is possible because they work with
abstract gates and gate errors.

\subsection{Rabi-frequency noise}


\begin{figure}[t]
\includegraphics[scale=0.7]{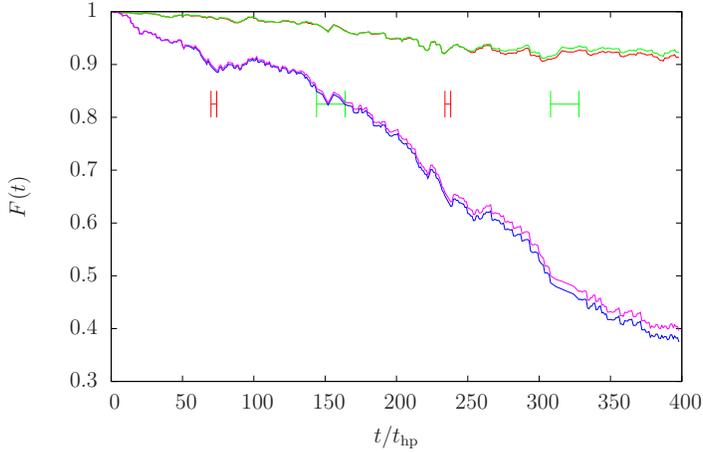}
\caption{Fidelity as a function of time for single runs in the presence/absence
of static Rabi frequency noise. Numerical simulations: without noise and exact
(red line); using the near resonant approximation (green line); with noise,
$\epsilon_R\xi =0.005$, and exact (blue line); using the near resonant 
approximation (pink line). Red and green vertical bars again indicate those
time intervals, where the oracle and the conditional reflections are applied.}
\label{f:RabiSingle}\end{figure}

In this section we study the effects of amplitude noise on the RF-field. In the
present approach this is equivalent to adding noise to the Rabi-frequency. As
in the previous case, we consider static and dynamic noise. In the case of 
static noise, the fidelity depends on only one random variable, the detuning of 
the Rabi-frequency: $\Omega \to \Omega_o + \epsilon_R \xi$, where $\xi$ is a 
random Gaussian variable with zero mean and unit variance. This detuning of the 
Rabi frequency changes at random for each repetition, while being kept fixed 
during the execution of the algorithm. In the case of dynamic noise, the 
detuning changes at random for each microwave pulse. As in the case of Larmor
frequency noise, we will first study the behavior of fidelity as a function of
time and only afterwards the final fidelity (for the execution of the whole
algorithm) as a function of the noise amplitude. We will find that the near
resonant approximation is accurate in the presence of Rabi frequency noise. In 
the study of the final fidelity $F_{\rm end}$, we therefore use that 
approximation.

Fig.~\ref{f:RabiSingle} shows the fidelity as a function of time for single 
runs of Grover's search algorithm. We compare simulations with and without 
static Rabi frequency noise in analogy to Fig.~\ref{f:LarmorSingle}, where we
showed a similar figure for the case of static Larmor frequency noise. Here,
a single run with static Rabi frequency noise, rather means that the simulation
is performed with a certain detuning: $\epsilon_R\xi= 0.005$. Comparing these 
results with those in Fig.~\ref{f:LarmorSingle}, we find that
a certain degree of fidelity loss requires a considerably larger amount of 
noise in the case of Larmor frequency noise ($\epsilon_L=0.02$). Furthermore,
we observe that the near resonant approximation is very
accurate in the present case of Rabi frequency detuning. 

\begin{figure}[t]
\includegraphics[scale=0.7]{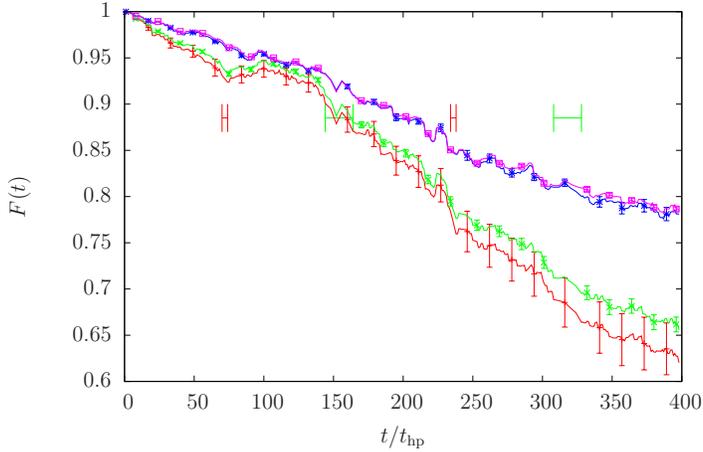}
\caption{Average fidelity as a function of time for static and dynamic Rabi 
frequency noise with amplitude $\epsilon_R=0.005$. Numerical simulations: 
Static noise, exact (red line); near resonant approximation (green line). 
Dynamic noise, exact (blue line); near resonant approximation (pink line).}
\label{f:RabiAverage}\end{figure}

Fig.~\ref{f:RabiAverage} shows the average fidelity for static (red line with 
error bars: exact numerical calculation; green line with error bars: near 
resonant approximation) and dynamic (blue line with error bars: exact numerical
calculation; pink line with error bars: near resonant approximation) Rabi 
frequency noise. The error bars represent the estimate of the statistical error
as explained in Eq.~(\ref{N:StatErr}). The exact calculations are performed 
with $n_{\rm rep}= 100$ repetitions, the near resonant approximations with
$n_{\rm rep}= 1000$. Note that the statistical uncertainty is quite large for
static noise whereas it is very small for dynamic noise. Comparing to 
Fig.~\ref{f:LarmorAverage} we observe much less pronounced differences between 
oracle and conditional reflection on the one hand and the Hadammard gates on 
the other. In both cases, static and random noise, the difference between exact
simulation and near resonant approximation is within the statistical error.

\subsubsection*{Static vs. Random Rabi frequency noise}

From the results for the fidelity as a function of time, we saw that the 
near resonant approximation works very well in the case of Rabi frequency 
noise. Since it provides an enormous speed-up of the simulations, we will use
it almost exclusively in the present section.

\begin{figure}[t]
\centerline{
\includegraphics[scale=0.8]{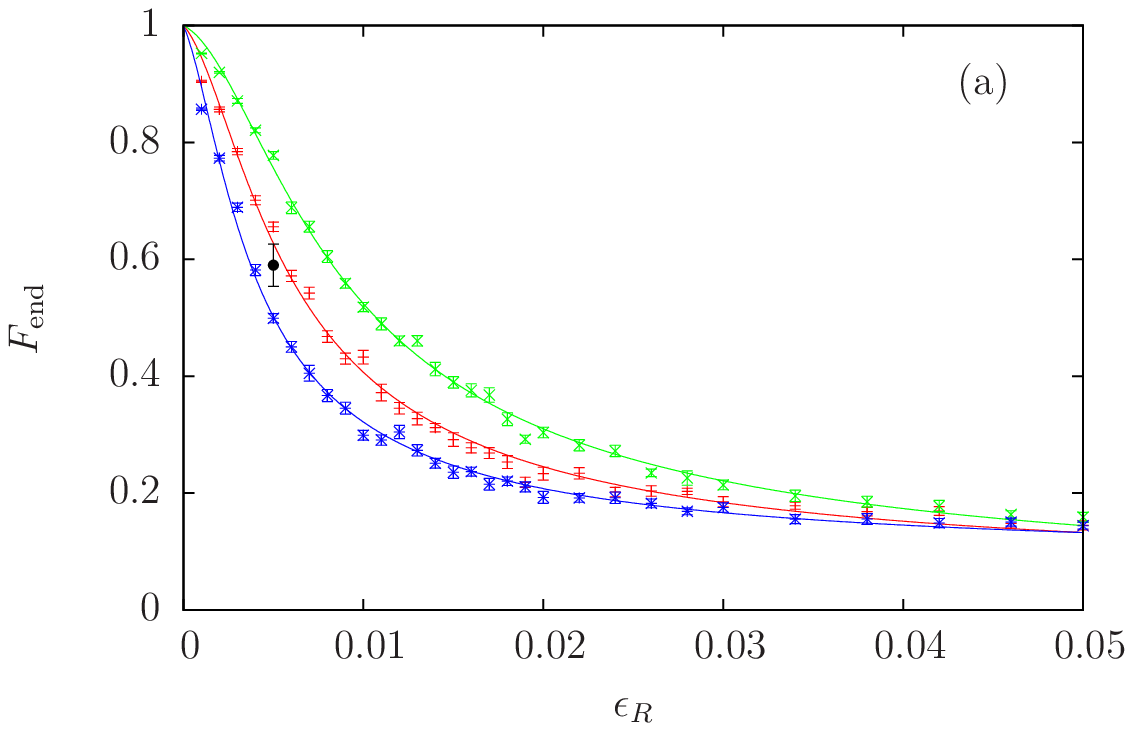}
\includegraphics[scale=0.8,clip=false,trim=1.6cm 0 0 0]{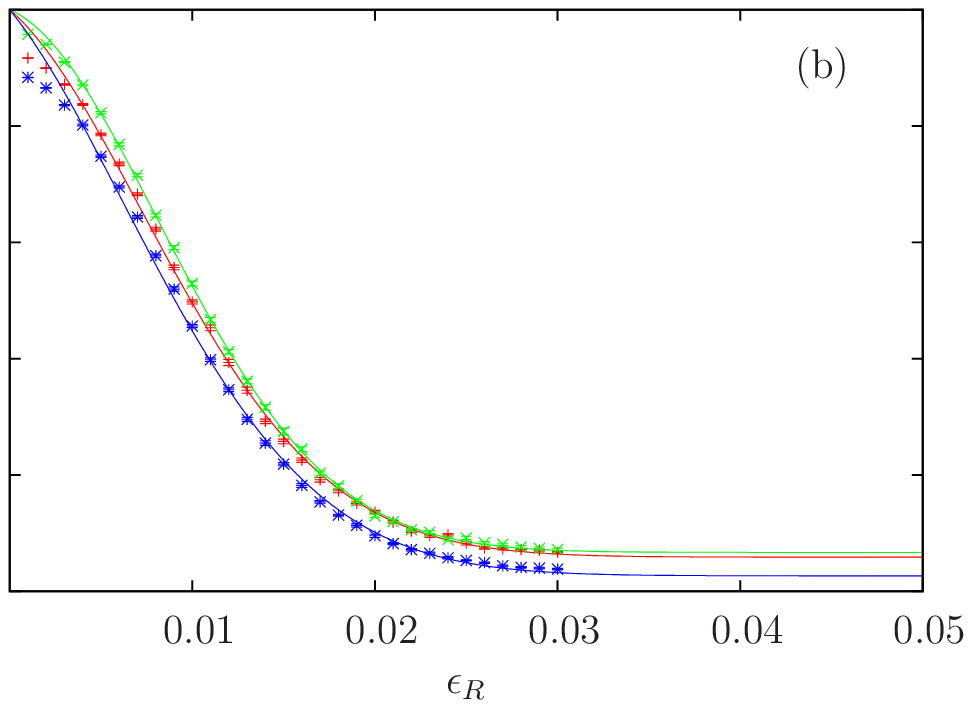}}
\caption{The fidelity $F_{\rm end}$ for static (a) and dynamic (b) Rabi 
frequency noise of strength $\epsilon_R$. Numerical simulations using the near 
resonant approximation: $4$ qubits, target state $0$ (red error bars); $4$ 
qubits, target state $8$ (green error bars); $5$ qubits, target state $27$ 
(blue error bars). The solid lines of the corresponding colour show the 
phenomenological fit with Eq.~(\ref{N:fit2}) (panel (a)) and Eq.~(\ref{N:fit})
(panel b) to the numerical data. The black error bar near the 
red curve shows one case, where the fidelity has been calculated exactly 
(static noise, $4$ qubits, target state $0$).}
\label{f:Rabi}\end{figure}

Figure~\ref{f:Rabi} shows the average fidelity $F_{\rm end}$ for the complete
algorithm as a function of the amplitude $\epsilon_R$ of the Rabi frequency
noise. In panel (a), we consider static noise, in panel (b), dynamic noise. We 
find a striking difference in the functional dependence of the fidelity between
the two cases. For dynamic noise (panel (b)) we expect the fit function 
Eq~(\ref{N:fit}) to work well. However in the case of static noise (panel (a))
we clearly need a different fit function to account for the apparently 
algebraic decay. In an attempt to maintain the general structure of the fit
function as much as possible, we choose the new fit function of the form
\begin{equation}
g(\epsilon) = g_{\rm bas} + \frac{1-g_{\rm bas}}
   {\sqrt{(\epsilon/\epsilon_2)^2 + \epsilon/\epsilon_1 + 1}} \; .
\label{N:fit2}\end{equation}
The following two tables contain the results of the fits for static and dynamic
Rabi frequency noise. 
\begin{itemize}
\item[(a)] Static noise: For the fit function $g(\epsilon)$ from 
Eq.~(\ref{N:fit2}) one obtains the following fit values for the three cases
considered:
\begin{equation}
\begin{array}{r|lll}
                      & g_{\rm bas} & \epsilon_2 & \epsilon_1\\
\hline
\text{4 qubits, O(0)} & 0.052(10) & 0.00429(23) & 0.0140(45) \\
\text{4 qubits, O(8)} & 0.0230(98) & 0.00633(24) & 0.032(10) \\
\text{5 qubits, O(27)} & 0.0798(69) & 0.00289(14) & 0.0064(14) \end{array}
\end{equation}

\item[(b)] Dynamic noise: Here, we use the same fit function $f(\epsilon)$ 
from Eq.~(\ref{N:fit}) as in the case of Larmor frequency noise, and we 
obtained the following fit values:
\begin{equation}
\begin{array}{r|lll}
                      & f_{\rm bas} & \epsilon_2 & \epsilon_1\\
\hline
\text{4 qubits, O(0)} & 0.0589(62) & 0.01437(44) & 0.0352(31) \\
\text{4 qubits, O(8)} & 0.0670(28) & 0.01325(16) & 0.0708(61) \\
\text{5 qubits, O(27)} & 0.0267(89) & 0.01499(72) & 0.0254(23) \end{array}
\end{equation}
\end{itemize}
The statistical uncertainties of the fit values are given in parenthesis.
They refer to the last two digits of the respective fit value.

Finally, we may again ask which kind of noise, static or random, is leads to a
larger loss of fidelity. While in Fig.~\ref{S:fig2} (Larmor frequency noise) it 
was the random noise, here in Fig.~\ref{f:Rabi} the situation is more or less
undecided. We note however that in the case of static noise (panal (a)) the
different cases lead to a considerable spreading between the results in 
contrast to random noise (panel (b)), where the corresponding curves are very
close together. We may also note that the $5$-qubit case in panel (a) shows 
initially the fastest drop in fidelity. It therefore has the correct tendency
if we believe that for more qubits and more Grover steps, the static errors 
should lead to lower fidelities~\cite{LoLi,Frahm04}.

Using the near resonant approximation, the numerical studies on Rabi frequency 
noise can be performed very efficiently. This would allow us to consider the
Grover search algorithm in a larger quantum register with more qubits. We would
need longer pulse sequences and more complex gate sequences for the 
implementation of the oracle. However, the main obstacle in considering more 
qubits are the unitary errors of our pulse sequence. Without additional error 
reduction ({\it e.g.} along the lines of Ref.~\cite{BerKam02}), the unitary 
errors alone would already reduce the final fidelity $F_{\rm end}$ so much that 
the effects of additional noise could not be analysed.

\section{\label{C} Conclusion}

We studied the effect of noise on the performance of Grover's search algorithm 
implemented on a nuclear spin quantum computer with four spins coupled via 
first- and second-neighbor Ising interactions. Starting from the ground state,
the quantum search algorithm attempts to build-up the target state in the 
quantum register, using the information obtained from inquieries of a quantum 
oracle. We used the fidelity to quantify the effect of noise on the final 
state, {\it i.e.} the result of the algorithm. We considered different types 
of noise: (i) Noise affecting the Larmor frequencies which would be due to
fluctuations in the static magnetic field, and (ii) noise affecting the Rabi
frequency, which would be due to fluctuations in the intensity of the 
radio-frequency pulses. We simulate the noise by randomly detuning the
respective frequencies from their nominal values. For both cases (i) and (ii)
we considered two types of noise, systematic and random errors. In the case of 
systematic errors, the resepective frequencies are randomly chosen at the beginning
of the algorithm and then kept fixed during the whole protocol. By contrast, in
the case of random errors, the respective frequencies are randomly chosen for
each radiofrequency pulse. 

We found that the threshold of this amplitude for still having good performance 
of the algorithm is $\epsilon_L\le 0.005$ for Larmor-frequency noise and 
$\epsilon_R\le 0.1$ for systematic error on Rabi-frequency .We also found that this algorithm 
is much more sensitive to systematic errors  than random errors which behave like an exponential and Gaussian  decay  respectively. One may say that this Gaussian behavior corresponds to the 
robustness of the algorithm. As we mentioned before, computing time increases very much with the number of qubits, making  the increasing of the number of Grover steps not feasible   within our model.

\end{document}